\def \SAIT #1 #2 {{\em Mem.\ Soc.\ Astron.\ It.\/} {\bf #1}, #2}
\def \MESS #1 #2 {{\em The Messenger\/} {\bf #1}, #2}
\def \ASTRNACH #1 #2 {{\em Astron. Nach.\/} {\bf #1}, #2}
\def \AAP #1 #2 {{\em Astron. Astrophys.\/} {\bf #1}, #2}
\def \AAL #1 #2 {{\em Astron. Astrophys. Lett.\/} {\bf #1}, L#2}
\def \AAR #1 #2 {{\em Astron. Astrophys. Rev.\/} {\bf #1}, #2}
\def \AAS #1 #2 {{\em Astron. Astrophys. Suppl. Ser.\/} {\bf #1}, #2}
\def \AJ #1 #2 {{\em Astron. J.\/} {\bf #1}, #2}
\def \ANNREV #1 #2 {{\em Ann. Rev. Astron. Astrophys.\/} {\bf #1}, #2}
\def \APJ #1 #2 {{\em Astrophys. J.\/} {\bf #1}, #2}
\def \APJL #1 #2 {{\em Astrophys. J. Lett.\/} {\bf #1}, L#2}
\def \APJS #1 #2 {{\em Astrophys. J. Suppl.\/} {\bf #1}, #2}
\def \APSS #1 #2 {{\em Astrophys. Space Sci.\/} {\bf #1}, #2}
\def \ASR #1 #2 {{\em Adv. Space Res.\/} {\bf #1}, #2}
\def \BAIC #1 #2 {{\em Bull. Astron. Inst. Czechosl.\/} {\bf #1}, #2}
\def \JSQRT #1 #2 {{\em J. Quant. Spectrosc. Radiat. Transfer\/} {\bf #1}, #2}
\def \MN #1 #2 {{\em Mon. Not. R. Astr. Soc.\/} {\bf #1}, #2}
\def \MEM #1 #2 {{\em Mem. R. Astr. Soc.\/} {\bf #1}, #2}
\def \PLR #1 #2 {{\em Phys. Lett. Rev.\/} {\bf #1}, #2}
\def \PASJ #1 #2 {{\em Publ. Astron. Soc. Japan\/} {\bf #1}, #2}
\def \PASP #1 #2 {{\em Publ. Astr. Soc. Pacific\/} {\bf #1}, #2}
\def \NAT #1 #2 {{\em Nature\/} {\bf #1}, #2}

\documentstyle{memsait}
\input epsf.sty
\begin{opening}
\title{THE AGILE CONTRIBUTION TO AGNs STUDIES}
\author{S.MEREGHETTI$^1$, A.PELLIZZONI$^1$,  M.TAVANI$^{1,2}$, G.BARBIELLINI$^3$, P.CARAVEO$^1$, A.MORSELLI$^4$,A.PERRINO$^4$,P.PICOZZA$^4$,P.SCHIAVON$^3$,S.SEVERONI$^4$, A.VACCHI$^3$}
\institute{$^1$Istituto di Fisica Cosmica "G.Occhialini", CNR, Milano, Italy\\
$^2$Astrophysics Laboratory, Columbia University, New York, USA \\
$^3$Dipartimento di Fisica, Universit\`a di Trieste e INFN, Italy \\
$^4$Dipartimento di Fisica,Universit\`a di Roma II, "Tor Vergata" e  INFN, Italy}
\date{} 
\end{opening}

\begin{document}

\oddpagefooter{}{}{} 
\evenpagefooter{}{}{} 
\ 
\bigskip

\begin{abstract}
The AGILE $\gamma$-ray telescope (Tavani et al. 1998) will have  a relatively uniform sensitivity
over a very wide field of view. 
This allows to simultaneously monitor several AGN's and to react quickly
with multiwavelength observations in the case of flaring activity.
The   large  cumulative sky coverage will also lead to
a significant increase in the number of $\gamma$-ray emitting
AGN's known at the end of the mission.
\end{abstract}

\section{The AGILE contribution to the study of Active Galactic Nuclei}

Contrary to  previous $\gamma$-ray instruments based on gas spark chambers (like COS B 
and EGRET), the  tracker of AGILE, using silicon strip detectors, does 
not need a separate triggering device. 
This allows to greatly increase the 
incidence angle of the accepted photons, resulting in a sensitivity that 
remains close to the on-axis value within a very large field of view. 
While EGRET at an angle of 30$^{\circ}$
has only $\sim$10\% of the on-axis effective area (Thompson et. al. 1993),
the AGILE  sensitivity is almost uniform within a radius of 
$\sim$ 50-60$^{\circ}$. 
For example, a single pointing toward 3C~273 (Fig.1)  
allows  to simultaneously observe and to look for significant flaring events 
in more than 20 sources 
(including  3C~279, Cen A, Mrk 421 and many other blazars).

A $\gamma$-ray outburst similar to that observed 
from the radio-loud quasar  PKS 0528+134
in March 1993 (Mukherjee et al. 1996) can easily be 
detected by AGILE with an integration  time of only one day (Fig.2).  
We plan to routinely search for such kind of events with  a quick look 
analysis of the data and to inform the scientific community in order to
react as soon as possible with coordinated observations.
 
Another advantage of the large  field of view is that 
the total exposure factor (cm$^2$ s)  for each direction of
the sky will be greater than that obtained by an
instrument with a smaller field of view.
We have estimated that, after a sequence of pointings equal to that
carried out in the EGRET phases 1 and 2 ($\sim$2 years of observations), the
average (over the whole sky) exposure factor reached by AGILE
would be greater by a factor $\sim$4. 
Since this reflects into a factor $\sim$2 lower limiting flux, 
we can roughly expect, assuming a $\gamma$-ray LogN-LogS with slope 3/2,
about three times more AGN's than the $\sim$70 discovered by EGRET. 
This is actually a lower limit since a further increase in sensitivity 
in confused regions at low galactic latitude  is obtained 
thanks to the  better angular resolution of AGILE (Morselli et al. 1998).

\begin{figure}
\epsfysize=6.cm 
\hspace{2.cm}\epsfbox{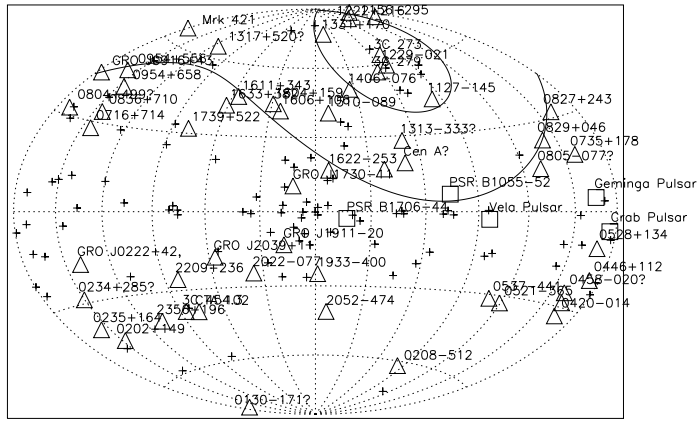} 
\caption[h]{Comparison between the AGILE and EGRET fields of view for a 
pointing toward   3C~279 and 3C~273
(the two regions correspond to radii of 25$^{\circ}$ and  60$^{\circ}$).}
\end{figure}

\begin{figure}
\epsfysize=6.cm 
\hspace{2.cm}\epsfbox{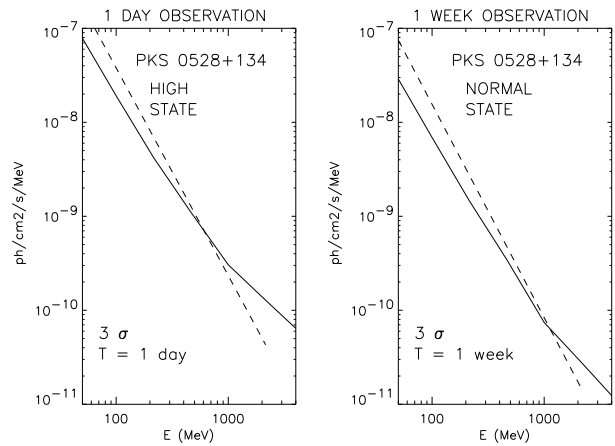} 
\caption[h]{Expected AGILE sensitivity (solid lines) for observations of 1 
day   and 1 week.}
\end{figure}


\end{document}